\documentclass[12pt]{iopart}
\usepackage[latin1]{inputenc}
\usepackage{a4wide}
\usepackage{psfig}
\usepackage{epsfig}
\usepackage{graphicx}
\usepackage{epsf}
\usepackage{amssymb}
\usepackage{verbatim}
\usepackage{fancyhdr}

\begin{document}

\title{Testing and design of a lens system for atom trapping and fluorescence detection}

\author{P Baranowski, J Zacks, G Hechenblaikner and C J Foot\address{University of Oxford, Clarendon Laboratory, Oxford OX1 3PU}}
\ead{P.Baranowski1@physics.ox.ac.uk}

\maketitle

\begin{abstract}
We present methods and results of the testing of an inexpensive home-made diffraction limited lens system, the design of which was proposed in \cite{i}
and which has since been used (with slight alterations) by several research groups (e.g.~\cite{3,6,9}). Our system will be used for both: focussing a collimated laser beam at a wavelength of $\lambda=830$~nm down to a narrow spot
and for collimating fluorescence light ($\lambda=780$~nm) emitted from rubidium atoms captured in this spot. Useful tests for lens systems include the use of ray tracing software~\cite{i+}, shear-wave interferometers~\cite{ii,iii}, the imaging of test charts~\cite{ii+} and
of polystyrene beads of a very small size~\cite{11,10}. We present these methods and show how conclusions can be drawn for the design under test.

\end{abstract}

\section{Introduction}
\begin{figure}
\begin{center}
\epsfig{file=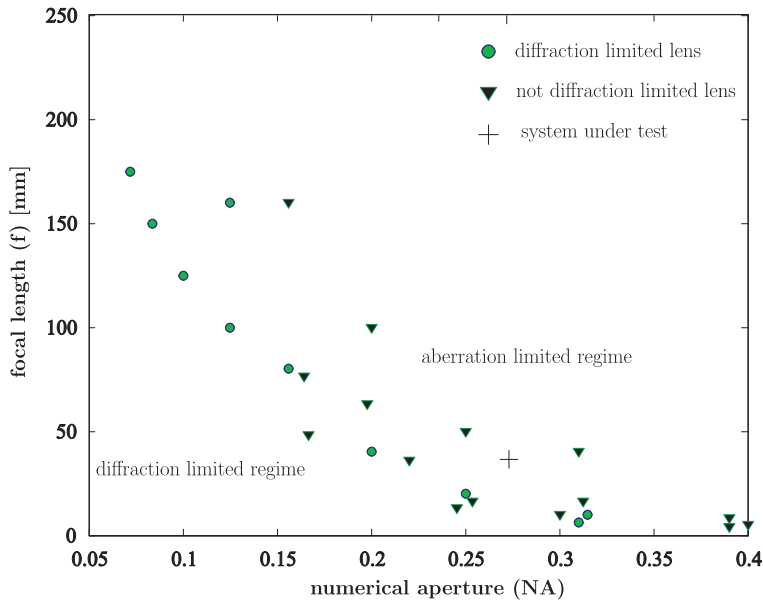, width=0.7\textwidth}
\caption{Performance of a selection of commercially available achromatic doublet lenses (Comar DQ series
and Newport VALUMAX{\textregistered} lenses). These data were gathered from manufacturers' specifications and ray tracing
results. The circles/ triangles denote diffraction limited/ aberration limited lenses. The '+' sign denotes our
objective lens. If a great NA is wanted only doublets from the catalogue assortment with very short f perform diffraction limited.
\label{1}}
\end{center}
\end{figure}

In some cases, the choice of appropriate lenses may decide upon
the success of an experiment. To keep lens aberrations small, experimenters often
choose corrected doublet lens systems, aspheric lenses or GRIN lenses which can have aberrations below the
diffraction limit. However, certain applications require a diffraction limited lens performance with
a high numerical aperture (NA) and a reasonable working distance (about $37$~mm in our case) at the same time, e.g. for picking up fluorescence
light from a weak source which is observed through the window of a vacuum cell. In many of such cases, commercially available lens systems are either very expensive or not available.
The lens system we discuss in the following (Fig.~\ref{3}c) was originally designed for the collimation of radiation from a point source, e.g. single ions emitting fluorescence light in a Paul
trap~\cite{vi}. In addition to the requirement of a high NA we also want to correct the spherical aberrations introduced by the plane silica window
through which the light source is usually observed. The setup currently under construction in our group even underscores the requirements of high resolution imaging. Cold neutral atoms can be trapped in optical dipole traps~\cite{iii+}. In dipole traps, laser light with a frequency far-detuned from the resonance of an electronic transition in
the atom interacts with the induced atomic dipole moment. The trapping potential can be calculated in terms of second-order time independent perturbation theory. The result is a shift in the ground states (``ac Stark shift'') which can
be exploited for trapping neutral atoms. In this context, two quantities are of general interest:
\begin{enumerate}
\item the dipole potential $U_{dip}(\mathbf{r})\propto\frac{\Gamma}{\Delta}I(\mathbf{r})$
\item the scattering rate $\Gamma_{sc}(\mathbf{r})\propto(\frac{\Gamma}{\Delta})^2I(\mathbf{r})$
\end{enumerate}
Here, $\Gamma$ denotes the dipole matrix element between the ground state and the excited state, $\Delta$ is the
detuning frequency and $I(\mathbf{r})$ the intensity of the electric field. Since the scattering rate scales as
$I/\Delta^2$, optical dipole traps usually use large detunings (about $50$~nm in our case) to keep the scattering rate as low as possible.
Consequently, high laser intensity is required to achieve sufficient trap depth. One aim of our experiment is to
produce a mini lattice made up of the spots of laser beams focussed through this objective under test. The spot size will finally
determine the intensity and hence, the trap depth. The lattice will be loaded with rubidium-87 atoms from a Bose-Einstein condensate and probed through exciting the $5\mbox{S}_{1/2}-5\mbox{P}_{3/2}$ (D2) transition. We slightly altered the design proposed in~\cite{i}
for optimum performance at the fluorescence wavelength of $780$~nm, which is the wavelength corresponding to the D2 transition.
However, since we use the same lens as well for focussing the dipole trap beam, an unvarying performance is requested as well for $830$~nm
light. The use of the objective for both, focussing laser beams and picking up fluorescence from the atoms make it a crucial part of our experiment and this requires a thorough testing. Fig.~\ref{1} shows the limitations of achromatic doublets for our needs.
For all statements about diffraction limited performance, we apply Rayleigh's Criterion as a check, that is
\begin{equation}
\Delta l=1.22\frac{f\lambda}{D}\mbox{,}
\end{equation}
where $\Delta l$ denotes the smallest resolvable distance, $f$ the effective focal length of the system, $\lambda$ the
wavelength of the light and $D$ the diameter of the aperture. In the case of our objective we have $f=37$~mm,
$D=25.4$~mm (NA=$0.27$) and hence $\Delta l=1.74~\mu$m ($\lambda=780$~nm).
\section{Designing and testing using ray tracing software}
A common way of designing and theoretically testing imaging systems is to use ray tracing software
which has become an indispensable complement to aberration theory. Built-in routines allow the designer to
optimize a system to specific needs by varying parameters such as the air gaps between lenses or the curvature
of surfaces. These programs specify rays to be traced through the system, compute the corresponding trajectories and also give some basic interpretation of the results. Usually the programs select a \emph{reference} ray which pass through some prescribed interior point in the system.
This ray is found by the program through tracing an arbitrary ray and then applying an iterative algorithm on it until the ray intersects the chosen point.
A common choice for the reference ray to intersect is the center of the aperture stop which makes it the \emph{chief ray}. Another type of ray is the \emph{ordinary} ray which starts from a certain point of the object in a prescribed direction. Both kinds of rays are called \emph{real} rays for
their trajectories are calculated to obey Snell's law exactly. In contrast, \emph{paraxial} rays are traced using the approximation $\sin(\alpha)\approx\alpha$
where $\alpha$ is the angle of the ray with the normal to the surface of a lens. This approximation is only good for rays traveling close to the optical axis.
Usually, the performance of an imaging system is in some way characterized by comparing the behavior of a number of ordinary rays with the reference ray at a certain point of
the system (e.g. the focus). The method of selecting appropriate ordinary rays is called \emph{ray aiming}. For most large aperture systems (NA$\ge0.1$) an aplanatic ray aiming
method is applied where the coordinates of the rays on the entrance pupil are chosen to be proportional to the direction cosines of ray angles. This method is equivalent
to aiming the rays at a sphere centered on the object point. The totality of rays traced through the system is called a fan of rays. Useful information about ray tracing principles and applications can be found in~\cite{5}.
\begin{figure}
\begin{center}
\epsfig{file=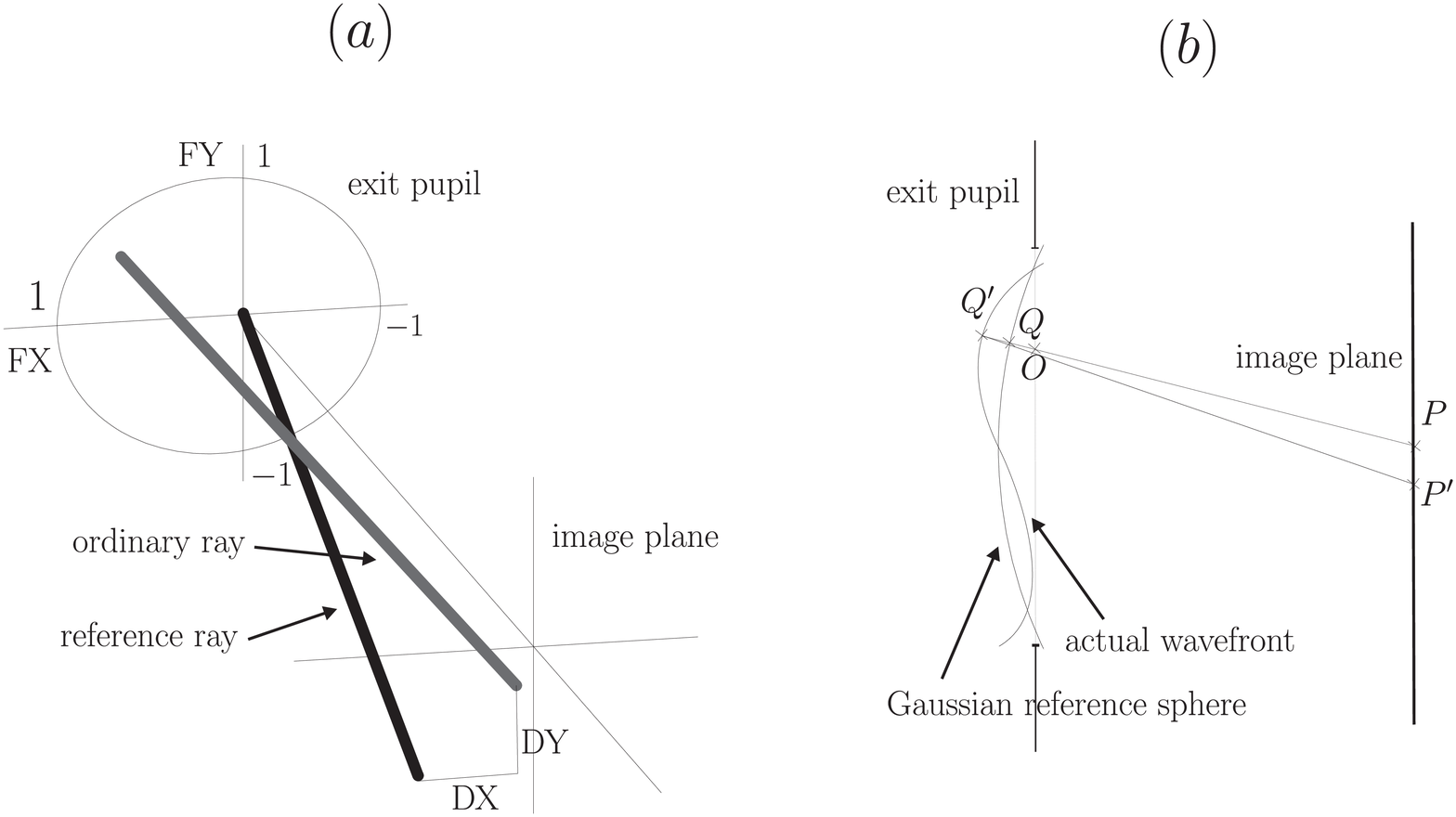, width=0.9\textwidth}
\caption{ways of quantifying aberrations: (a) ray aberration vector (DX,DY) e.g. depicted in ray intercept curves as a function of the fractional pupil coordinates (FX,FY) (b) wave aberration $\overline{QQ'}$ e.g. measured
in shear-wave interferometers in units of fractions of the wavelength. P is the image point of an aberration-free system (center of Gaussian reference sphere). Q and Q' are the points of intersection of an actual ray with the Gaussian reference sphere and with
the actual wavefront respectively. P' is the actual image point.\label{2}}
\end{center}
\end{figure}

\subsection{Interpreting ray data}
\begin{figure}
\begin{center}
\epsfig{file=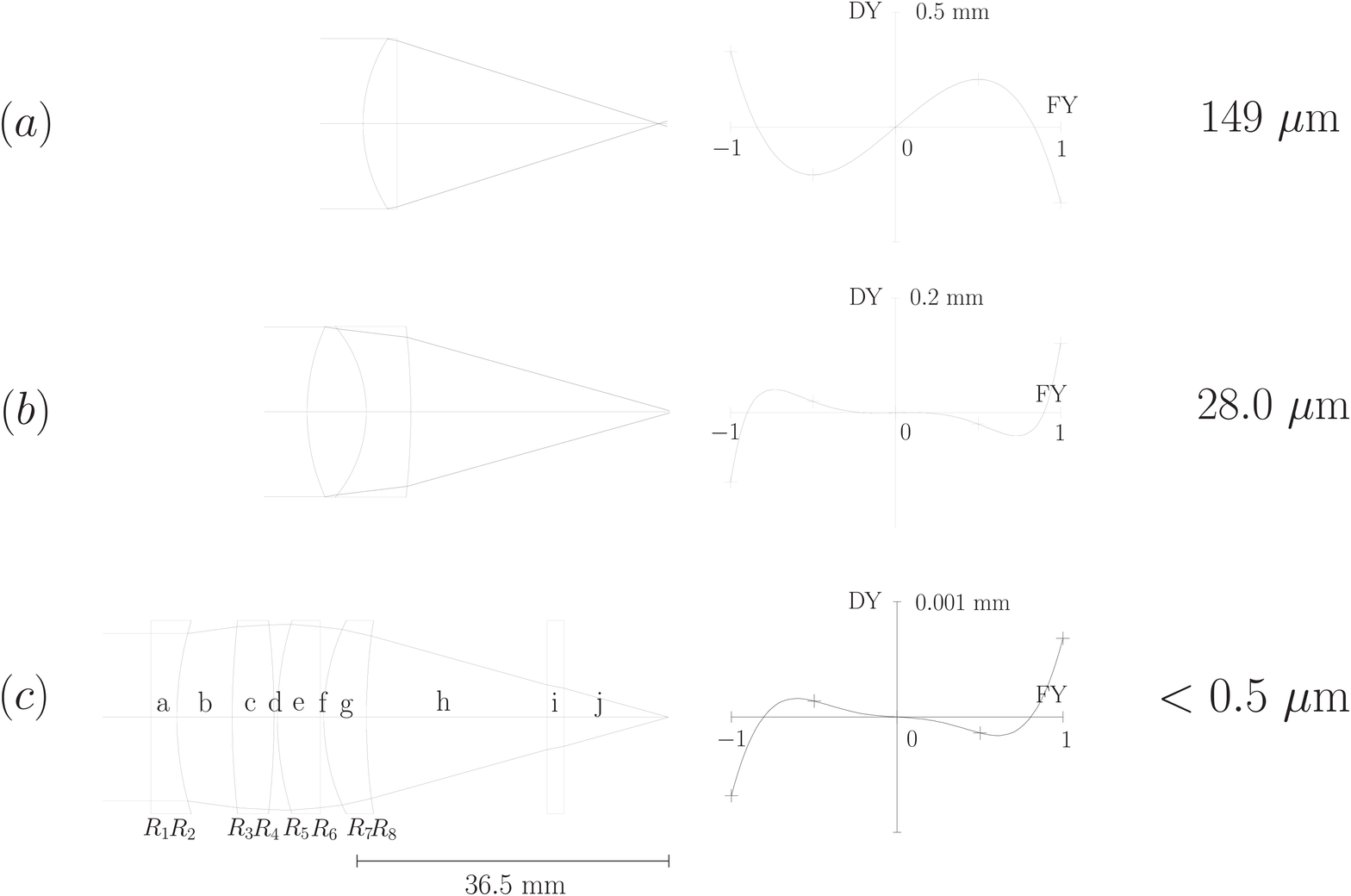, width=0.9\textwidth}
\caption{design, meridional ray intercept curves and rms spot sizes for (a) plano-convex lens, (b) cemented doublet lens, (c) four lens system.
The simulations (OSLO) were run for a Gaussian beam with $20$~mm $1/e^2$-diameter for light of $780$~nm wavelength. The numerical aperture for a,b and c is $\approx 0.27$, the diffraction
limit is $1.74~\mu$m, hence only (c) performs diffraction limited.\label{3}}
\end{center}
\end{figure}

Ray intercept curves and optical path difference (OPD) evaluations are used to measure aberrations of an optical system. The respective principles are explained in Fig.~\ref{2}. Intercept curves
are graphical plots of ray fans (Fig.~\ref{2}a). In Fig.~\ref{3} we show such curves for meridional fans. The characteristics of intercept curves allow designers not only to deduce the amount of
aberrations in a system, but also to get first ideas about the type of aberration contributing most. A discussion of
how the types of individual aberration alter the shape of the curve is beyond the scope of this
paper. However, another very useful quantity that can be deduced from this ray analysis is the root mean square (RMS) spot size radius $\sigma_r$ of an imaging system. This quantity represents
the theoretical spot size when diffraction effects are neglected and is defined as:

\begin{equation}
\sigma_r=(\frac{1}{W}\sum_{i=1}^nw_i[(DX_i-<x>)^2+(DY_i-<y>)^2])^{1/2}
\end{equation}

The sum extends over the $n$ rays traced through the system, $DX_i$ and $DY_i$ are the two components of the aberration
vector which are measured relative to the point of intersection of the reference ray, $<x>$ and $<y>$ are the centroids of the spot

\begin{equation}
<x>=\frac{1}{W}\sum_{i=1}^nw_iDX_i\hspace{2cm} <y>=\frac{1}{W}\sum_{i=1}^nw_iDY_i~\mbox{,}
\end{equation}

$w_i$ is a number representing a weight of the ray considered
and $W=\sum w_i$. $\sigma_r$ can be compared directly to the diffraction limit of a system and hence determine whether
it is diffraction limited or not.\\
As another tool, the OPD (Fig.~\ref{2}b) can also be specified if the image plane is at infinity and can hence be applied e.g. for the investigation of collimated beams.
A minimum standard invented for high-quality optical performance is that the wavefront is distorted or deformed by spherical aberration less than $\lambda/4$ peak-to-valley (P-V), with $\lambda$ being the
wavelength of yellow-green light (Rayleigh's quarter wavelength rule). This can be more generally expressed as a maximum allowed RMS departure of the wavefront of about $\lambda/14$ \cite{iv}.

\section{Lens data}
The lens system discussed here consists of three standard lenses and one meniscus lens (Fig.~\ref{3}c).
The glass is BK7 for all lenses. They were obtained from~\cite{12}. The system was optimized for our needs ($\lambda=780$~nm and a silica window of $2$~mm thickness)
using the optimization routine of the ray-tracing program.
During this procedure, the radii of curvature of the lens surfaces were fixed to catalog values and the
lens distances were used as variables. This allows the program to correct for the squared sum of the spherical aberrations up to
seventh order and third order coma and astigmatism. Once the appropriate values were obtained through the iterations,
it was also verified that the performance
will be comparably good when manufacturer's tolerances are taken into consideration. These tolerances correspond mainly to the
precision to which the lenses were mounted within a brass tube and kept separated by thin aluminium spacer rings. The data for the lens surface curvatures are (in mm):
\emph{$R_1$} flat, \emph{$R_2$}=$39.08$, \emph{$R_3$}=$103.29$, \emph{$R_4$}=$-103.29$,\emph{$R_5$}=$39.08$,
\emph{$R_6$} flat, \emph{$R_7$}=$26.00$, \emph{$R_8$}=$78.16$. The widths of the lenses and air gaps are (in mm):
\emph{a}=$3.08$, \emph{b}=$7.24$, \emph{c}=$4.97$, \emph{d}=$0.40$, \emph{e}=$5.12$, \emph{f}=$0.40$, \emph{g}=$5.07$, \emph{h}=$21.00$, \emph{i}~(silica)=$2.00$, \emph{j}=$12.73$.
All lenses have a diameter of $25.4$~mm.

\section{Experimental testing of lens systems}

\subsection{Shear wave interferometer}
\begin{figure}[h]
\begin{center}
\epsfig{file=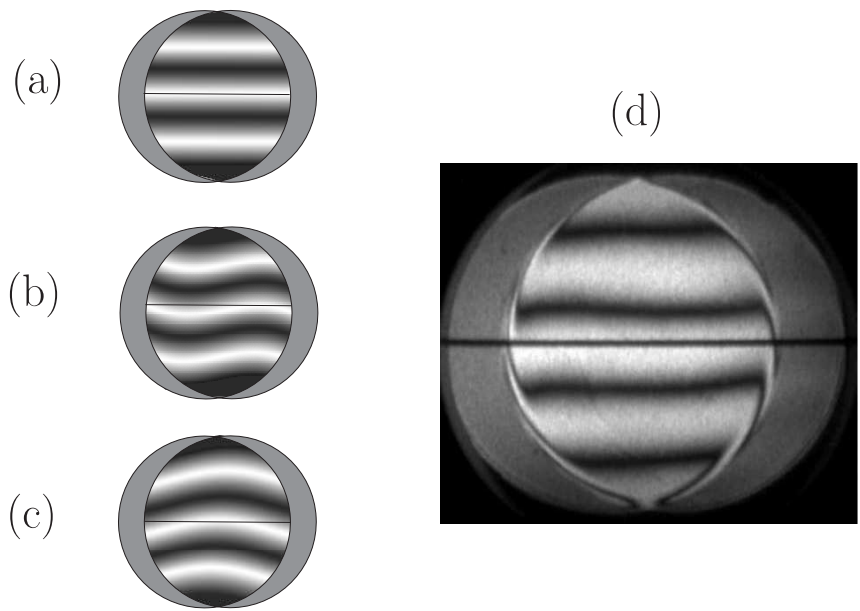,width=0.7\textwidth}
\caption{Study of wavefront aberration. Left: model sketches for collimated beams with (a) no aberrations, (b) $\lambda/4$ spherical aberrations and (c) $\lambda/4$ comar.
Right: (d) an actual wavefront measurement for the lens proposed in \cite{i} (Fig.~\ref{8}d courtesy of Urban \& Fischer-Verlag)}
\label{8}
\end{center}
\end{figure}

Shear wave interferometers are useful tools for testing the collimation of a laser beam. The principle of operation is
fairly simple: the front and rear surface of a high quality optical flat are used for reflecting the beam at an angle
of $45°$. The two surfaces are not quite parallel but slightly wedged to produce a graded path difference between
the front and back surface reflections of the shear plate. When the two reflected beams overlap on a screen, they produce a linear fringe
pattern. For a perfectly collimated beam, the interference pattern will consist of a series of equally spaced straight fringes.
A noncollimated beam which is reasonably close to collimation will produce the same pattern, but rotated w.r.t.
the fringes from collimated light. The radius of curvature can be deduced from the angle of rotation. Wavefront aberration
will distort the straight lines of the pattern in characteristic ways~(Fig.~\ref{8}a-c). Hence, from looking at the fringe pattern, one can roughly deduce
 the magnitude and the kind of main aberration in an optical system~\cite{ii,iii}. Further practical information can be found in~\cite{4}. For the case of our lens system, it is convenient to
collimate light from a point source (e.g. a $1~\mu$m pinhole onto which a laser beam is focussed) and analyze it using the interferometer.
Following this procedure, the P-V wavefront aberration for the objective lens in~\cite{i} was estimated to be less than $\lambda/4$ (Fig.~\ref{8}d).

\subsection{Test chart imaging}
\begin{figure}[h]
\begin{center}
\epsfig{file=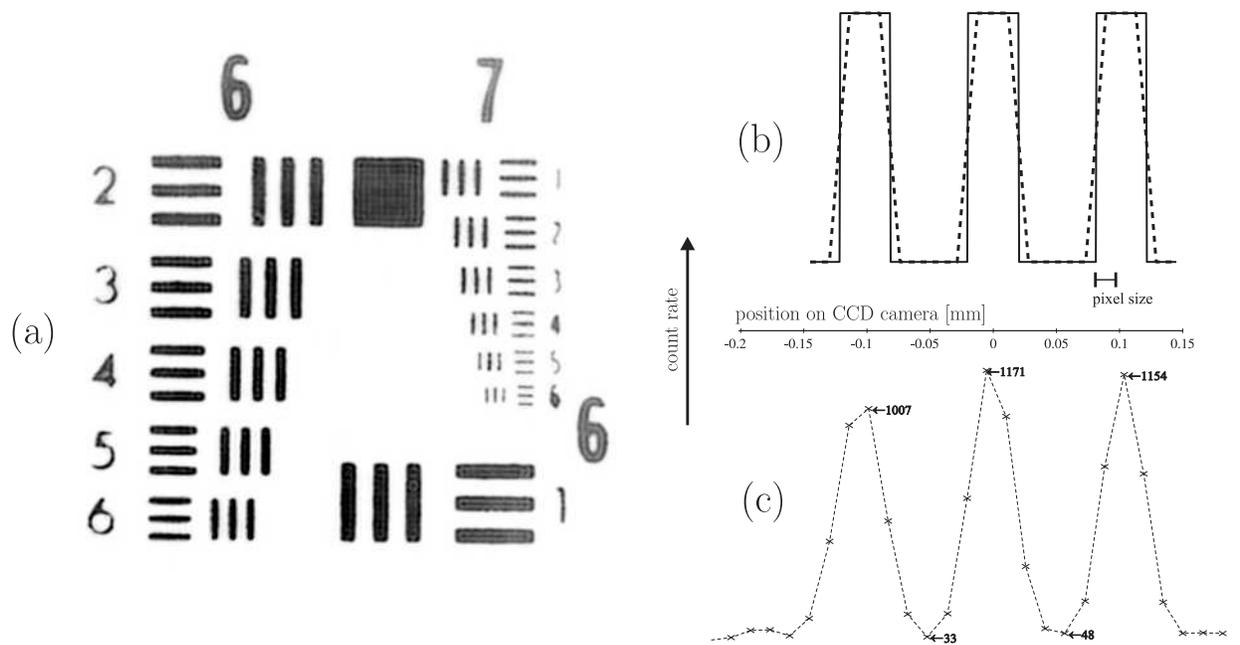,width=1.0\textwidth}
\caption{(a) Inverted picture of the finest region of USAF R70 test chart obtained with our objective. The smallest line separation is $4.4~\mu$m. Important information
on the performance was obtained from measuring the contrast of the finest structure of the chart. (b) Solid line: sharp steps of the finest structure.
Dashed line: convolution with the pixel function (single step of $16~\mu$m width).
The dashed line shows how the sharp structure gets washed-out due to the finite CCD pixel size. We used a magnification of $\approx~25$ to get useful
data.(c) an actual measurement (points are linked for clearness). The dashed line of Fig.~\ref{4}b is what would be obtained by averaging over many measurements.\label{4}}
\end{center}
\end{figure}

A very common way of quantitatively testing the performance of a lens is to image an appropriate test chart~\cite{ii+}.
The \emph{USAF 1951 Test Target} was invented for testing the resolution of a lens and it is one of the most commonly used charts.
Furthermore, a variety of more specialised charts are available, e.g. by using \emph{Dot Distortion Targets},
one can determine the amount of distortion from an array of precisely placed dots in a regular array. \emph{The Sector Star Target} consists of equally sized bar and space segments
and was designed to test for astigmatism. However, a couple of performance features beyond resolution can be deduced from imaging a resolution chart onto a CCD camera~\cite{viii}. Some methods are presented
in the following. For the image shown in Fig.~\ref{4}a
we illuminated the chart with weak laser light (few $\mu$W of laser power, $\lambda=780$~nm) and imaged it to a plane $\approx25\times\mbox{EFL}$ away from the back focal plane, where
EFL denotes the effective focal length of the system ($37$~mm). Even though the image distance is considerably larger than the object distance, the
performance of the objective is better when it is used for collimation instead of imaging. Hence, all results given in this examination should be considered as lower limits of the lens performance.
For critical applications it is recommended to image the collimated
light with low NA diffraction limited optics.

\subsubsection{Astigmatism}
An oblique parallel ray bundle incident on a lens will suffer from astigmatism, i.e. the focal lengths for meridional and sagittal rays in this
bundle will be different. Our system performs without pronounced astigmatism for ray angles (or, equivalently, tilts of the objective w.r.t. the optical axis) $<0.25~$deg.
This number was deduced from measuring the different contrasts for the finest horizontal and vertical stripes of the test chart against tilts of the objective lens
(Fig.~\ref{6}a). The contrast was always defined as max/min counts in measurements such as Fig.~\ref{4}c. Note that by using this definition, the contrast becomes sensitive to the
counts in the minima. To obtain comparable data, it is therefore essential to keep the background constant within a test series.
The observation of astigmatism on test charts can be effectively used for the alignment of the system.

\subsubsection{Field of view}
One of the requirements for the objective lens was specified by the designer as a field of view of about $1$~mm~\cite{i}. It is important
to have a sufficiently large field of view, since the position of trapped atoms in a vacuum cell cannot be perfectly specified and might additionally extend over a greater range.
Furthermore, when the objective is used for focussing laser beams to create dipole traps, a great field of view provides a uniform trapping potential over a convenient range.
Ray tracing simulations affirm a diffraction
limited performance over a range of about $1$~mm (compare Fig.~\ref{6}d). We evaluated the contrast of the finest stripes while the objective was translated
horizontally/ vertically w.r.t. the optical axis. From this measurement, we could deduce a field of view of about $0.4$~mm (Fig.~\ref{6}b). This is also in agreement
with ray tracing predictions for the actual experimental way of testing the lens (see above).

\subsubsection{Focal tolerance}
\begin{figure}[h]
\begin{center}
\epsfig{file=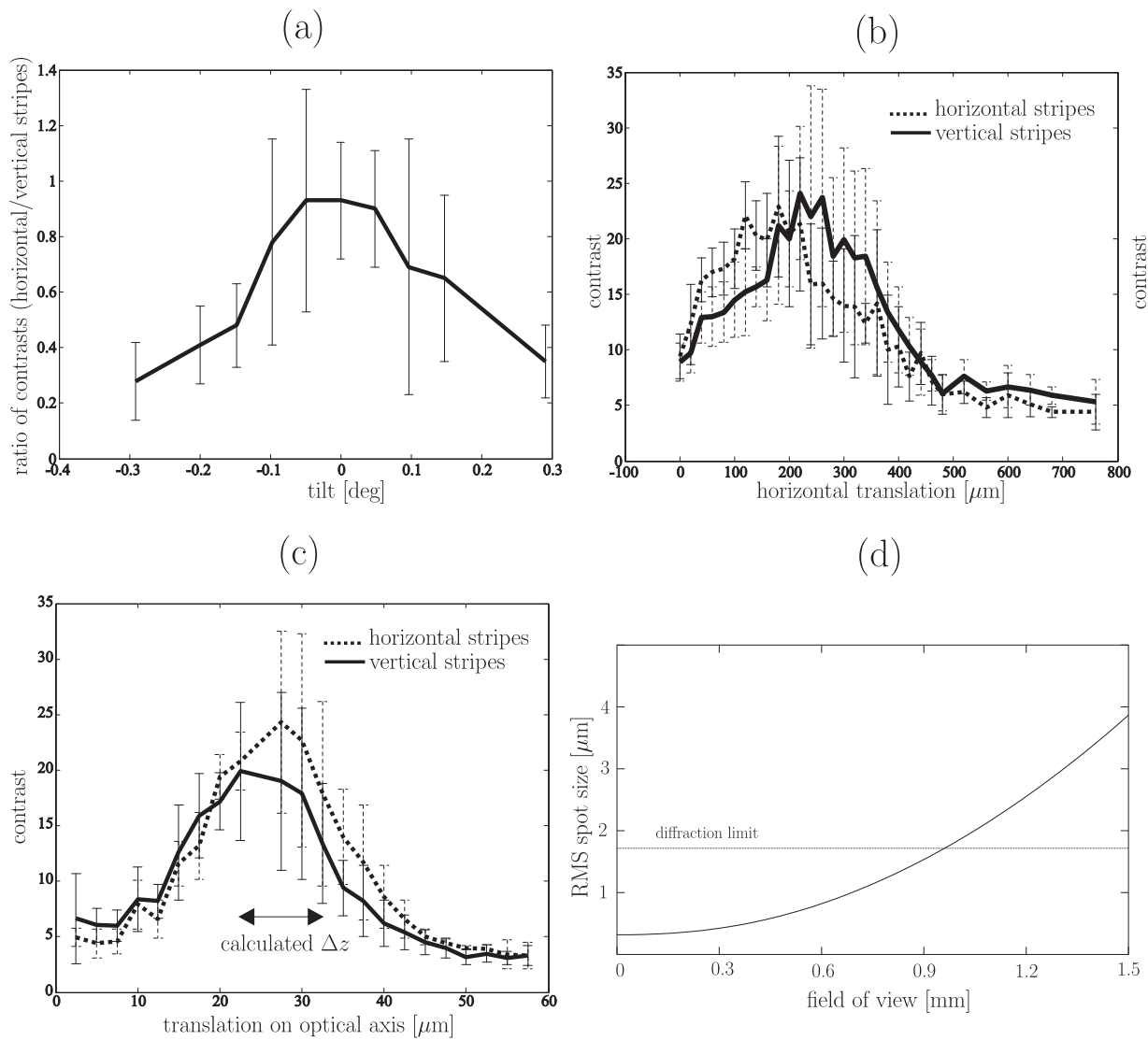,width=1\textwidth}
\caption{measuring the contrast of the finest horizontal/ vertical stripes of the test chart while tilting/ translating the objective.
From this, we determined the
(a) amount of astigmatism, (b) field of view and (c) focal tolerance.
The contrasts were deduced from plots such as Fig.~\ref{4}c. The error bars take into account the spread of the measured
counts. The respective points are linked for clearness (solid line, dashed line). (d) Spot sizes for points at the verge of the respective field of view.
Diffraction limited performance can be expected for a field of view up to about $1$~mm (ray-tracing program).}
\label{6}
\end{center}
\end{figure}

From evaluating the diffraction integral for a spherical monochromatic wave emerging from a circular aperture in the proximity of the focal point, one can deduce an expression for the focal tolerance $\Delta z$~\cite{vii}. This distance is defined
by the convention, that a loss of $\approx20\%$ in the intensity of the focal point along the axis is permissible. It follows:
\begin{equation}
\Delta z\approx\left(\frac{f}{a}\right)^2\lambda
\label{c}
\end{equation}
where a denotes the radius of the aperture of the lens. For our system, this yields $\Delta z=10.4~\mu$m. We have roughly verified this by measuring the contrast of the finest stripes of the test chart (Fig.~\ref{6}c).
The similar behavior for vertical and horizontal stripes in Fig.~\ref{6}b) and c) is a good indicator for the centring of the system.

\subsubsection{Resolution}
\begin{figure}[h]
\begin{center}
\epsfig{file=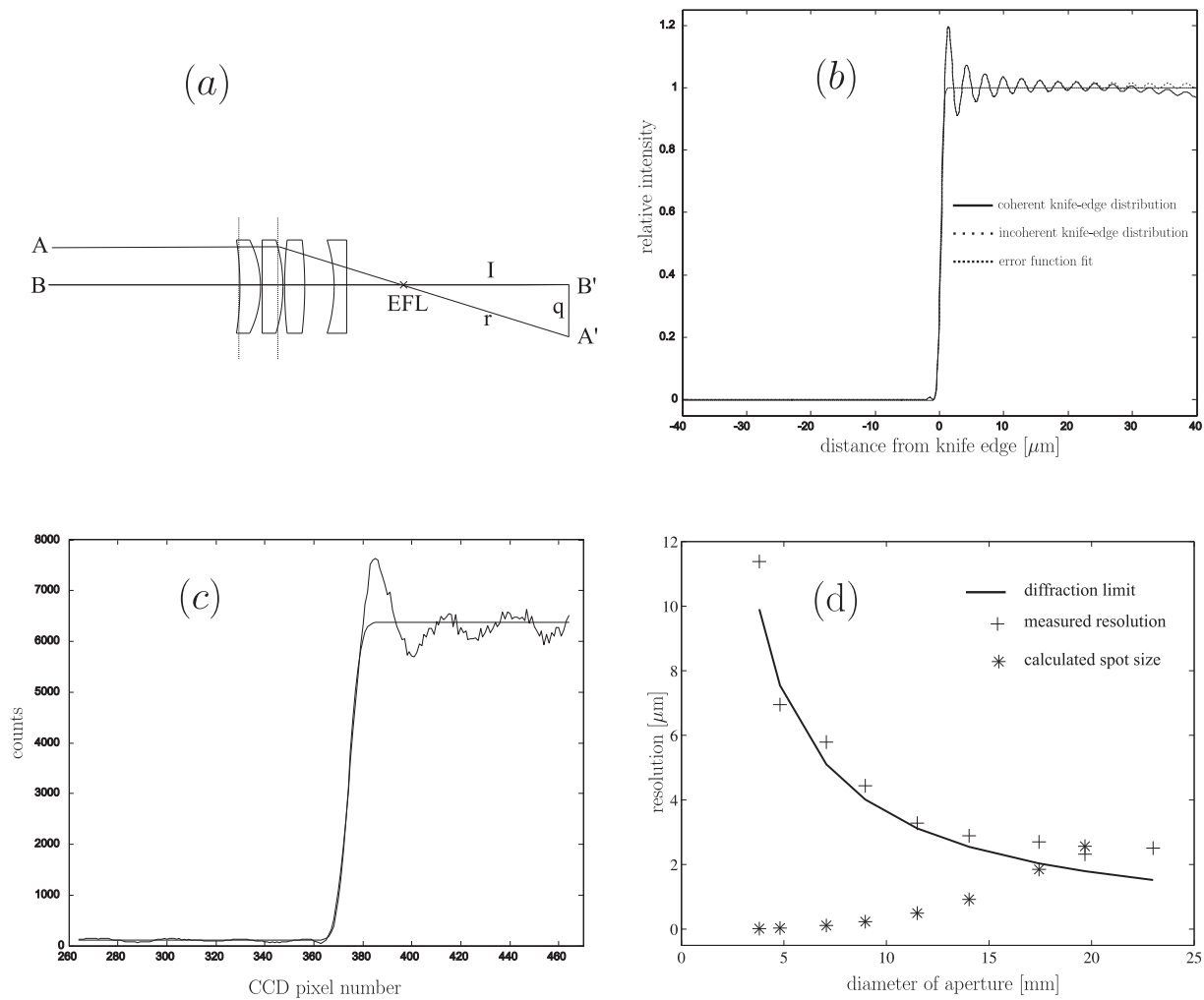,width=1.0\textwidth}
\caption{(a) derivation of the relative phase difference between A' and B': $\phi\approx\frac{\pi q^2}{2I}$, (b) numerical integration of Eq.~\ref{a} with and without phases for a diffraction limit
of $1.74~\mu$m and error function fit,
(c) an actual experimental measurement for a diffraction limit of $10~\mu$m with the objective under test and error function fit (magnification 25.4), (d) resolution measurements for different apertures. The
solid line depicts Rayleigh's diffraction limit (Eq.~\ref{1}). $\lambda=780$~nm for all sketches.}
\label{5}
\end{center}
\end{figure}

As shown in Fig.~\ref{4}, the resolution is obviously considerably better than $4.4~\mu$m. In an attempt to specify the resolution more precisely,
 we analyzed images of the edge of a very broad stripe ($\approx 100~\mu$m) of the test chart.
The edge was considered as a sharp knife edge. When laser light is used, the actual shape of the image will be the
convolution of a step function with a coherent transfer function. For a purely diffraction limited system, the transfer function is the
electric field E(x,y) as obtained from Fraunhofer diffraction at a circular aperture:

\begin{equation}
\mbox{E}(x,y)=\mbox{E}_0\frac{2\mbox{J}_1(\theta)}{\theta}\qquad\mbox{where}\qquad\theta=\frac{2\pi aq}{\lambda I}\mbox{.}
\label{h}
\end{equation}
Here, $x$ and $y$ are the Cartesian coordinates in the image plane and $q^2=x^2+y^2$. $\mbox{E}_0$ is the amplitude of the electrical field,
$\mbox{J}_1(\theta)$ is the Bessel function of first order, $a$ is the radius of the aperture and $I$ is the image distance minus the EFL (Fig.~\ref{5}a).
Moreover, we have to consider the phases. The convolution integral yields:

\begin{equation}
\mbox{KED}(X)=[\int_{-\infty}^{X}dx'~\int_{-\infty}^{\infty}dy'~\mbox{E}(x',y')\cos\phi(x',y')]^2
\label{a}
\end{equation}

where $\phi(x,y)$ is the relative phase given to a good approximation as $\phi=\frac{\pi q^2}{\lambda I}$~\cite{7} (compare Fig.~\ref{5}a).
Finally, $\mbox{KED}(X)$ (\emph{'Knife-Edge Distribution'}) expresses the actual one-dimensional intensity of the knife edge image as a function of the scanning coordinate $X$ on
the $x$-axis. The numerical evaluation of Eq.~\ref{a} is shown in Fig.~\ref{5}b. It can be seen that the phase factors
have no practical relevance. An experimental measurement for a small aperture is shown in Fig.~\ref{5}c.\\
For practical purposes, the KED measured in an experiment can usually be well fitted to the error function:
\begin{equation}
erf(X/w)=\frac{2}{\sqrt{\pi}}\int_0^{X/w}e^{-t^2}dt\mbox{.}
\end{equation}
Here, $w$ corresponds to the $1/e$-waist of a Gaussian distribution.
Rayleigh's criterion (Eq.~\ref{1}) defines the resolution as the distance between the peak and the first minimum of Eq.~\ref{h} and hence $\Delta l$ it is
determined by $\theta$. After the convolution, $\Delta l$ will be encoded in the width of the transition at the edge and therefore there is a linear relation
between $\Delta l$ and $w$.
We integrate Eq.~\ref{a} numerically for a known resolution (i.e. for a certain $\theta$) and fit an error function to the result.
Thereby we obtain the following simple formula for the resolution as a function of the fitting parameter $w$:

\begin{equation}
\Delta l=3.38\frac{w}{\mbox{magnification}}\mbox{.}\label{b}
\end{equation}

Note however, that this formula is only good for systems which perform in reasonable proximity of the diffraction limit. For
highly aberrated systems, a considerable amount of intensity will be outside the Airy disk. In addition to broadening the transition in KED($X$), this alters the shape of the distribution.
Consequently Eq.~\ref{b} delivers erroneous results in such cases. It should rather be used for verifying that a system performs diffraction limited instead of calculating the actual resolution of a system
in which aberrations prevail.
In the same sense, $\Delta l$ can be determined from the periodicity of the intensity oscillation on the top edge of KED($X$) (compare Fig.~\ref{5}b).
The theoretical treatment yields a linear relation between the periodicity and the resolution:

\begin{equation}
\Delta l=\frac{\Lambda}{1.78\cdot\mbox{magnification}}
\end{equation}

where $\Lambda$ denotes the periodicity of the oscillation. This was verified within an error of $5$\% in an actual measurement (Fig.~\ref{5}c). In Fig.~\ref{5}d we show the result of resolution measurements (obtained by using Eq.~\ref{b}) for different apertures (an iris was attached close to the
aperture stop). The solid line depicts Rayleigh's diffraction limit (Eq.~\ref{1}).

\subsection{Measuring the PSF directly}
\begin{figure}[h]
\begin{center}
\epsfig{file=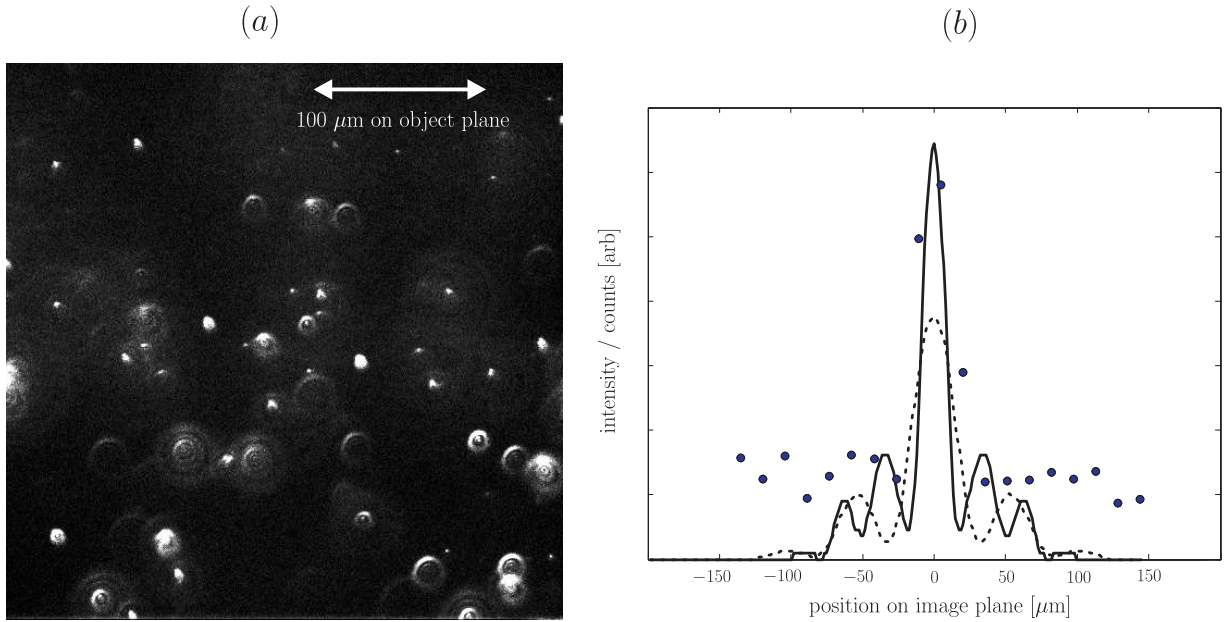,width=1\textwidth}
\caption{(a) Imaging of light scatter from small polystyrene beads, (b) calculated normalised Point Spread Functions for $488$~nm light (solid line) and $780$~nm (dashed line) (OSLO). The
measurement (filled circles) was done with light of wavelength of $488~$nm. The CCD pixel size is $16~\mu$m, the magnification is $25.4$ for both sketches}
\label{9}
\end{center}
\end{figure}

From looking at the Point Spread Function (PSF) for a lens, one can deduce many important properties of the system including
resolution and amount of aberrations. It expresses the intensity distribution of the image of a mathematical point source.
The PSF of a diffraction limited system is usually given by the Airy function, i.e. the square of $E(x,y)$ in
Eq.~\ref{h}. A way to measure the PSF directly is to image particles that are smaller than the diffraction limit. One can either
image light scatter from Polystyrene beads or fluorescence light emitted from convenient sources~\cite{11}.
Adequate beads are commercially available in a great variety of sizes. We tested both, fluorescence and light-scatter particles.
In this report we will only focus on the latter. We chose a sample with a diameter of $1~\mu$m~\cite{10}, diluted the concentrate by
a factor of $1000$ with water in a glass cuvette and illuminated the suspension with laser light. The beam intersected the cuvette slightly off the EFL to produce an image
with a magnification of about 25. Due to the finite beam diameter, not only beads within the focal tolerance scatter light, but also particles
that are further off-focus. Consequently, the image on the CCD consists of sharp points for in-focus beads and blurred circles for off-focus beads.
Although, for our purposes, the design of the system was optimized for a wavelength of $780~$nm, ray tracing simulations predict a diffraction limited performance
for wavelengths down to about $400$~nm. By using the Argon-Ion $488$~nm line for the illumination of the beads, we tested the performance for a wavelength somewhat
shifted from the design wavelength. For a quantitative evaluation, we compare the CCD counts of the image
for an in-focus bead with the PSF obtained from ray tracing simulation. For the determination of the width of the main peak of a measured PSF, one has
to consider the finite pixel width of the CCD camera used, which washes-out the original image depending on the position of the pixels with respect to the
structure of the image. For the PSF shown in Fig.~\ref{9}, in the worst configuration, the error in the measured width can be as high as $46\%$. Within that error the measured PSF width is in
good agreement with the predictions (solid line). For reliable statements one must either use a sufficient great magnification, a small pixel size or to take several measurements.

\section{Conclusion}
We have assembled a diffraction limited objective lens with an EFL of $37$~mm and an active aperture of $20~$mm following the proposal in~\cite{i}.
The system will be used in an experiment for both: focussing a laser beam of $830$~nm laser light to create a tight optical dipole trap and for collimating fluorescence light at $780$~nm collected from atoms confined
in this trap.
We have conducted a number of straightforward tests on the system which showed agreement with predictions of ray tracing software and hence, a diffraction limited
performance. The methods described here can easily be adopted for testing this or other lens systems.

\section*{Acknowledgements}
We acknowledge stimulating discussions with Geoffrey Brooker and Herbert Crepaz, and would like to thank Richard Berry's group for providing us with test beads as well as useful information and help.
This work was supported by the EPSRC.

\end{document}